# Collision of Ionization Waves in Long Discharge Tubes

*A I Shishpanov , D O Ivanov and S A Kalinin*

*Saint Petersburg State University, Universitetskaya emb. 7/9, 198510 Saint Petersburg, Russia*

e-mail: a.shishpanov@spbu.ru

(Dated: December 28, 2018)

This research focuses on interaction of ionization waves (IW) generated in long discharge tubes (1 m length and 1.5 cm in diameter) filled with neon or argon under pressure about 1 Torr and, in particular, on simplest case of such interference which is collision of the opposite directed IW of positive polarity. The waves were triggered from two electrodes of the tube by simultaneous application of the same high voltage pulse and propagated toward the tube center until their collision. Space and time stability of the collisions was achieved via decreasing time delay of the initial breakdowns which trigger IW by employment of both memory effect and photo-desorption of wall surface electrons. Significant memory effect occurred for the pulse repetition rate over 100Hz. Multiple IW transition from the electrode forms specific one-electrode discharge; as a result the problem of IW collision was transformed into the observation of two opposite one-electrode discharges interaction. Patterns of IW interfering were registered via deviation of their kinematic characteristics (*x-t* diagrams) from ones for non-collided wave. This method shows that two ionization waves travel toward each other with decreasing speeds and subsequent degradation of the emission. An area near the tube center where IWs emission was not detected was considered as the IW suppression zone. The zone width depends on voltage amplitude, gas type and pressure.

## I. INTRODUCTION

The most advanced researches regarded generation and development of ionization waves (IW) during electrical breakdown in long discharge tubes in wide range of gas pressures, but just a few works considered interaction between IW. This direction of experimental and theoretical studies of ionization waves may exhibit some new properties of them and confirm the validity of accepted models. The present work focuses on experimental research of a simplest case of the IW interaction which is the mutual attenuation (suppression) of two waves running to the same point each from the opposite electrode distanced enough to ensure the independence of the wave starting.

Ionization wave can be briefly described as a wave of electric field, bell-shaped in space and time, which propagates from high voltage electrode over the discharge gap producing gas ionization [1]. The ability of IW to conserve its shape during the propagation makes it similar to a solitary wave. IW appears as a result of electric field non-uniformity specific for the tube which length exceeds its diameter (long discharge tube). The most common examples are fluorescent lamps [2] and tubes of gas discharge lasers [3]. In such geometry, the electric field before breakdown rapidly decreases from electrodes toward the tube center, and becomes zero at distances in some tube diameters [4]. In this case ionization processes occur generally near the high voltage electrode, and extend to the discharge volume in a wave-like manner with velocity depending generally on applied voltage. According to the wave velocity IW can be classified as slow or fast. If IW speed ranges from 105 up to 108 cm/s such wave is considered slow. This wave initiates the breakdowns in gases at low pressures (p ~ 1Torr) and under comparably low voltages (U < 10kV). Fast waves with velocities 108 – 1010 cm/s are usually observed at high pressures (p >100Torr) and voltages up to 300kV [9]. After IW approaches the opposite electrode the return wave starts from it and moves in the reverse direction with the velocity exceeding the initial IW velocity by an order of magnitude. In some breakdown conditions [5, 6] several IWs must pass over the gap before glow or arc discharges appear. The most studied parameter of IW is the propagation velocity which depends on value and sign of applied voltage, discharge gap geometry (tube radius), gas type, pressure and initial electron concentration.

Ionization waves appear during not only breakdowns in long tubes. For example, most recent works concerning IW consider this phenomenon in the context of plasma jet investigation. In this case IW plays the role of the discharge precursor making the ionized gas track for the plasma jet development (plasma bullet) [7]. In spite of the great difference in the conditions of IW existence (from low pressure breakdown to atmospheric pressure plasma jet) the basic properties of the phenomenon such as the wave shape, patterns of the propagation velocity and attenuation are conserved. But behavior of IWs under their interaction has yet to be fully understood. Some papers listed below as the authors know contain the results of observation of the interaction at different conditions.

Interaction of ionization waves guiding plasma jet was observed in paper [8]. The authors of the work studied the plasma jet generation in neon and neon/xenon mixtures, and investigated processes of splitting and merging of ionization wave guiding the plasma jet. The discharge was ignited in a system consisted of two linear tubes both connected to a central ring-shaped tube along the same longitudinal axis being the ring diameter. The standard system consisted of tip-shape inner electrode and ring external electrode was used to produce the plasma jet. The breakdown appeared after loading the inner electrode with -25 kV amplitude voltage pulse (25 ns rise time) while the external electrode was connected to grounded metal box which covered the whole system. Both electrodes were assembled in the inlet tube. It was found that IW splits in first branching point into two equal fronts that propagated in the half-rings with lower speeds than the origin wave did. Speed decreasing was a result of a lower electric field strength. In second branching point the two waves merged in one front with higher electric field. The new IW had an increased speed and plasma density compared with IW in half-rings, but these parameters were lower than these of the wave in the inlet tube. From the diagrams and image presented in the paper it can be assumed that when two IW merge their fronts become distorted so that the waves can exist neither in the outlet point nor in the ring-tube. This result shows that IWs collision with each other leads to suppression of both fronts which can be explained by superposition of oppositely directed axial components of electric fields in central point between two approaching IWs. One might assume that if the ring-tube did not have the outlet channel, the waves would decompose due to the compensation of electric fields in the area behind the IW fronts and subsequent decreasing in ionization.

Work [9] shows an opposite situation of the wave interaction when two IW move in the same direction with different speeds. This was the case when breakdown was initiated by a voltage pulse with negative polarity (rise time ~10 ns) triggering the first wave start from the cathode within the induction time and moving in non-ionized gas at the speed about $10^7$ cm/s. Formation time of the cathode spot exceeded the induction time of IW, so the spot appeared during IW movement which led to sharp emission of electrons into the plasma. Arising space charge generated the second IW which moved in previously ionized gas at a speed one order of magnitude higher than that of the first wave and overtook it in some point of the tube. After the waves superposition the new IW appeared and moved in the remaining part of the tube with intermediate speed. Similar phenomenon was observed in [10] when breakdown pulses were applied to discharge tube electrodes with rate $2 \cdot 10^5$ Hz, so the pulse period was less than the time needed for IW to overcome the inter-electrode gap. In such experiment the next wave generated before than the previous one approached the opposite electrode and moved in pre-ionized gas at a higher speed. This IW motion can repeat several times prior the discharge ignition. As a result each subsequent IW approached the previous wave and overlapped with it forming a new IW.

The above examples of IW interactions show that the results of overlapping several IW could be diverse depending on directions of the interacted waves. If the waves propagate in the same direction they could form new IW after the interaction. But in case of oppositely directed waves the new IW cannot appear on axis along which the interaction took place. Obviously these results should be tied with electric fields superposition principle during overlapping of ionization waves having nonlinear character. Thus it is possible to assume the existence of complex interference of IW resulting for example in the waves amplification or in their suppression as it takes place in case of electromagnetic waves. The present article explores the case of such interference which is the mutual attenuation of two IW moving to the same point from two opposite electrodes (collision of two ionization waves).

## II. EXPERIMENTAL

The experiment was conducted in a sealed-off 1 m long and 1.5 cm diameter tubes containing neon at pressure of 1 - 4 Torr and argon at pressure of 4 Torr. The tubes contained two aluminum electrodes having the form of hollow cylinders 2 cm in length and 5 mm in diameter with ceramic collars at the front edges.

To observe the collision of two ionization waves we tried to satisfy several conditions. First of all the IWs must be similar in amplitudes of electric field strength and velocities thus the result of the interference will be easier to analyze. Another condition required the independence of IWs to be collided. As a result at least two sources of independent identical IWs were needed. To achieve that we used a simplest scheme in which the same voltage pulse is applied simultaneously to both electrodes of the tube initiating two oppositely directed IWs. The pulses had a rectangular shape 2.1 – 3.5 kV in amplitude, rate of voltage 2kV/μs and duration 0.25 – 10 ms. Frequency of the pulses sequence was ranged from 5 up to 100 Hz. All voltage parameters were controlled by a high-voltage probe and a digital oscilloscope (Tektronix TDS 240). In such method two identical IW started simultaneously within their local induction times and traveled toward each other. It is clear that at such tube connection the steady state discharge can not exist but ionization waves easily appear. The main condition of maintaining the wave motion is the presence of high potential at the electrode from which it starts. If the pulse duration is less than the time needed for IW traveling over the discharge gap its motion will break with voltage.

Under this property of IW a specific type of gas discharge may exist – one electrode discharge (OED). This discharge is excited by application of the series of high voltage pulse to only one electrode of the tube while another one remains free. OED in long tubes at low gas pressure was described in [11, 26]. The discharge develops in three stages: 1) initial breakdown; 2) IW propagation; 3) development of reverse ionization wave. The first and second stages are well known for breakdowns in long tubes and described in numerous works (see [6, 12, 13] and references therein) but the reverse IW was recently found and published for the first time in [11]. In contrast to the standard IW the generation of this wave appears at trailing edge of the voltage pulse but it also propagates from active electrode and demonstrates all properties of IW. The mechanism of this phenomenon could be explained as follow. Development of breakdown IW at leading edge of the voltage pulse is accompanied by charging of the tube wall. The charge value equals to the charge which is transmitted through the active electrode during IW motion. In other words, the current from IW front to the electrode charges the wall inner surface like a plate of capacitor and loops via displacement current to the surrounding grounded elements of the set up. After the voltage pulse breaking the electrode potential becomes zero while the wall surface has the high potential of the surface charge. This voltage difference leads to the repeat of the initial breakdown from the nearest wall surface section to the electrode triggering ionization wave which discharge the tube wall during own propagation.

In common case we studied the overlapping of two OEDs that start at both electrodes of the tube simultaneously. Due to the low intensity of the optical signal the reverse waves were not recorded in experiment and we supposed that whole emission of the discharge was composed from signals of initial breakdown and IW propagated. Thus we can consider OED as a source of single ionization waves that appear with the same rate as the discharge pulses. It is worth to stress that the classical two-electrode breakdown which studied in works cited above is invalid as such source due to the complexity of optical signals appertaining now not only to the breakdown IW but also to the return IW (moving from low voltage electrode) and to the steady state discharge. In some conditions several traveling of direct and return waves need to complete the breakdown. These properties will complicate the analysis of the resultant optical signal after IW interaction.

Diagram at figure1 shows the details of electrical circuit of the set up. The high voltage pulses were generated by power source 2 with maximum output dc voltage 4kV and high voltage switch 3 operated by series of TTL pulses from digital generator 11. The ballast resistor 4 restricted slowly varying components of the current while the forming resistor 5 ensured the trailing edge of the voltage pulse. It was included to discharge the tube after the pulse termination. The voltage is applied to both electrodes of the discharge tube 1 via the common connection point. Compensated voltage probe 6 allows to control the voltage time running in common point of the electrodes connection. Photomultipliers 7 and 8 detected emission signals transmitted by optical fibers 9. The optical and voltage signals were recorded by digital oscilloscope 10 Tektronix TDS240 with the band width 100 MHz which was sufficient for registration IW signals having characteristic time about 100 ns. Data processing and storage as well as the operation of the set up were performed from PC via corresponding software.

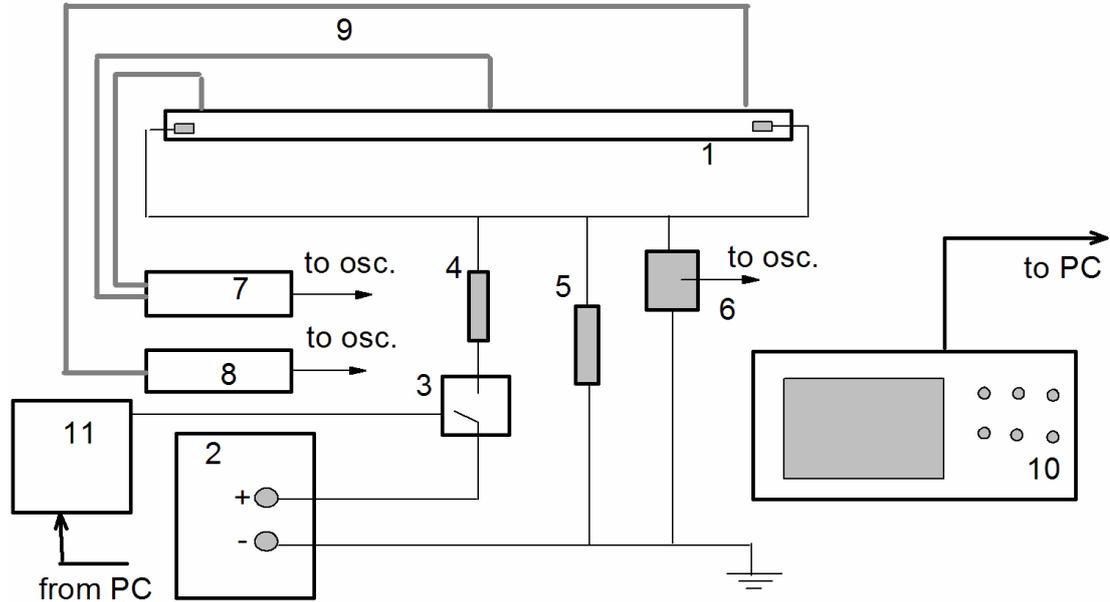

**Figure 1** Block–scheme of the set-up. 1. discharge tube; 2. power source; 3. high voltage commutator switch; 4. ballast resistor 3.9 kΩ; 5. forming resistor 1.0 MΩ; 6. compensated high voltage probe; 7 and 8 photomultipliers; 9. optical fibers; 10. digital oscilloscope; 11. operating generator

Another problem which should be solved is the synchronization of two IWs of the opposite one-electrode discharges. Indeed after voltage application the discharge starts with the delay time $t_d$ consisting of the delay time of the initial breakdown $t_{di}$ [13] and the IW induction time $t_{ind}$ of IW, that is $t_d = t_{di} + t_{ind}$. Therefore the IW initiation time may vary within $t_d$, and the synchronization problem reduces to restriction of this value. The largest part of $t_{di}$ as well as of $t_d$ is the statistical time $t_s$ which is determined by appearance of effective electrons [14]. The time $t_s$ is a stochastic value determined by the appearance rate of initial electrons $Y$ and the probability of breakdown $W$, so: $\overline{t_s} = 1/YW$ [15]. To reduce $t_s$ we used two methods of increasing the value $Y$. First of them is photo-emission under the effect of visible light [16] emitted by two diode lasers λ = 406 nm. The lasers illuminate the tube wall at 3 cm from the electrodes. According to the results of the works [13, 17] such method may produce electrons releasing with the rate over $2·10^5$ s$^{-1}$ in 1 Torr neon, and decreasing in average $t_s$ from about 1ms up to ≈1μs. It was also found that laser illumination leads to the IW movement stabilization manifested by a decreased jitter observed in the moments when the IW reaches certain points of the tube.

Another method consisted in the development of memory effect [13, 18] by increasing the voltage pulse repetition rate. Repeated passages of IW through the discharge gap lead to accumulation of residual metastable species which become a sources of electrons for the initial breakdown. Enlargement in initial electrons concentration ensures the decreasing of $\overline{t_s}$. The release of electrons during de-excitation of metastable states at the tube wall and the Auger effect at the electrode surface are the most probable mechanisms of the memory effect [19]. Simultaneous application of both methods gives the value of $\bar{t}_d$ less than 100 ns.

According to the idea of the experiment, the discharge tube was not covered with a metal screen like in the work [20]. It is clear that the metal cage disturbs the IW electric field. But in order to exclude the influence of grounded metal elements of the set-up, the tube was distanced at least 20 cm from any metal items. Mounted elements were made of a thin wooden panels, and arrangement of the equipment was conserved during the whole experiment.

Interaction of two opposite IW was observed via registration of the waves in more than 30 points along the tube axis. Since the tube was non-shielded, and application of capacity sensor was impossible, we detected the IW movement via its optical emission which were transmitted by three optical fibers. Two of them were fixed and placed at 3 cm from the electrodes while the last one was movable and could be displaced along the tube. The fibers were metal-less and approached the front edge to the tube wall as near as possible. The experimental test shows that such action does not influence the IW propagation. Light from the discharge was transmitted via the optical fibers to the two photomultipliers (PMT). The movable fiber and one fixed fiber which registered the start of IW near the active electrode were both connected to the one PMT while one more fixed fiber was connected to the second PMT to detect the moment of the opposite IW appearance. The PMT signals were recorded by digital oscilloscope Tektronix TDS 240 and then processed using a computer.

To register the interaction of two opposite IW we analyzed *x-t* diagrams [9, 21] for both waves. A standard method of *x-t* diagrams requires measurement of locations passed by the same IW along they trajectory. This procedure would require a lot of optical fibers and a set of recording equipment. Using the breakdown stabilization methods described above we achieved a sufficient repeatability in generation of both IW, and following the practice described in the paper [9] we measured the locations of IW for different voltage pulses as a function of time like for the wave from only one pulse. In order to exclude errors which can be attributed to the small jitter in $t_d$ the each PMT oscillogram corresponded to an average of 128 measured records. So the measurement procedure was as follows. The moveable optical fiber was placed in a certain point along the tube; after that the oscilloscope recorded the PMT signal 128 times and gave an averaged picture of the optical signal containing the two picks: one from fixed and another from moveable fibers. The picks show the most probable time points when IW has the same coordinates as the fibers. Knowing the position of the displaced fiber with respect to another it is easy to plot an average *x-t* diagram for IW. Interday recordings of the same *x-t* diagrams show their acceptable repeatability. Due to the impossibility to manufacture two exactly similar electrodes the *x-t* characteristics of the one-electrode discharges ignited from opposite electrodes of the same tube will differ from each other. This effect seems to be bound with the state of the electrode surface which almost cannot be controlled in the sealed off tubes.

**Experimental procedure and data processing**

The experiment was conducted in two stages. At the first stage we investigated one-electrode discharge starting at either electrode separately in order to observe the intrinsic motion of IW. In this experiment only one electrode was active while another electrode was free. The *x-t* diagrams show the patterns of the one-electrode discharge in the tube under several voltages. To ensure the tube wall electric charge and state of the electrode surface specific for the long-term one-electrode discharge, series of pulses were ignited as a preparative discharge with duration 10 – 20 min prior to the measurements. The *x-t* diagrams analysis consisted in finding the dependence of IW velocity $V_{iw}$ on the coordinate *x*. If ionization wave velocity changes slightly during the propagation time *x-t* diagram of such wave will be close to straight line. IW of this type has named stationary ionization waves [9, 21, 22]. In contrast to stationary waves ionization waves observed in the presented work demonstrated marked non-uniform motion. The *x-t* curves were fitted with polynomial regressions. First derivative of the fitted function will give IW velocity versus time, and plotting of this value as a function *x* will correspond to the desired dependency $V_{iw}(x)$.

The interaction of the opposite waves was studied at the second stage of the experiment when the same high voltage pulse was applied to both electrodes simultaneously. The *x-t* characteristics were recorded for either of waves and analyzed as described above. Run of the plots $x(t)$ and $V_{iw}(x)$ clearly shows the waves interaction area which dimension can be then easily measured.

### III. RESULTS

**1. One-electrode discharge.** Graphs at figure 2 show typical optical emission signals from one-electrode discharge transmitted from two points: (1) fixed point near the electrode and (2) changeable point (20 cm from the electrode at the discussed graph). The optical signal of the wave has the form of the peak with the width at half height ranging from 80 ns up to 140 ns in 4 Torr Argon, and 100 – 180 ns – in 1 Torr Neon. This parameter is tightly bound to the IW velocity as it is higher than the pick is narrower. Such picks compose optical emission of the one-electrode discharge only. No any subsequent signals from either return or reflected waves are detected which makes the observed picture easier for the analysis. Transition of electrical current in such discharge is possible in the form of short pulses only, but their multiple repetition leads to uniformly glowing plasma column already at voltage pulse rate less than 100 Hz, so the one-electrode discharge can be observed with the naked eye as a uniform glow column filling the whole tube (figure 7).

The examples of the one-electrode discharge *x-t* diagrams are shown in the figure 3 for several voltage amplitudes. The curves demonstrate generally non-uniform motion of IW. For lower voltage the deceleration becomes stronger (curve 1), while at the increasing voltage the wave motion tends to the stationary type when the *x-t-* characteristic tends to the straight line (curve 3).

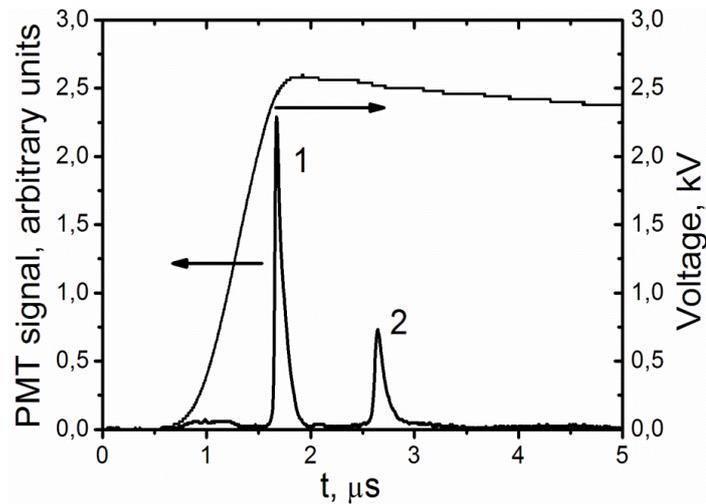

**Figure2** One-electrode discharge optical emission (1) near active electrode, (2) at 20 cm from it; run of the voltage at the electrode after power applying. Ar, $p = 4$ Torr, amplitude of voltage 3.0 kV

The observed deceleration with the voltage decreasing leads to the situation when IW attenuates before its coming to the opposite tube end thus one-electrode discharge occupies not the whole tube. At minimum voltage the discharge is observed just near the active electrode. It was found that the discharge length cannot be less than 4 cm. Similar result was found for IW of both polarities in [22]. The authors of this work stressed that IW forms at some distance from the active electrode which is the minimal value of coordinate x of the kinematic characteristic. Plots on figure 4a demonstrate the patterns of the discharge plasma boundary (i.e. the length of one-electrode discharge) displacement with the increase of voltage amplitude showing nearly linear dependency. The higher the voltage the longer the discharge. We also observed the IW attenuation in case of partial tube occupation with the discharge in the form of *x-t* diagrams recorded for different positions $x_0$ of the plasma boundary. OED length ($x_0$) was controlled by amplitude of the applied voltage.

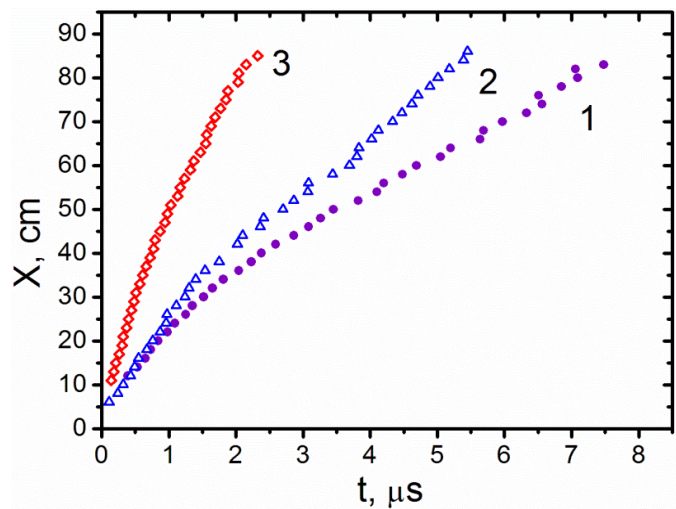

**Figure 3** Examples of *x-t* diagrams of one-electrode discharge (1)Ar, $p = 4$ Torr, voltage amplitude 3.0 kV; (2) Ne, $p = 1$ Torr, voltage amplitude 2.1 kV; (3) Ne, $p = 1$Torr, $U = 3.1$ kV , $f = 100$ Hz

The plots (fig. 4b) correspond to the three values of $x_0$: 25 cm, 51 cm and 70 cm counted from the active electrode. The data scattering near $x_0$ was induced by instability of IW under strong attenuation which was manifested in much more random appearance of the wave in the given point of the tube. That is natural enough because the decrease in IW amplitude (maximum value of electric field) leads to decrease in it ionization ability and therefore to unstable propagation. All this curves approach the horizontal lines corresponding to the stop of IW, while the rate of the curve transition to the horizontal line shows the rate of IW attenuation. For the shorter discharges the attenuation is much higher.

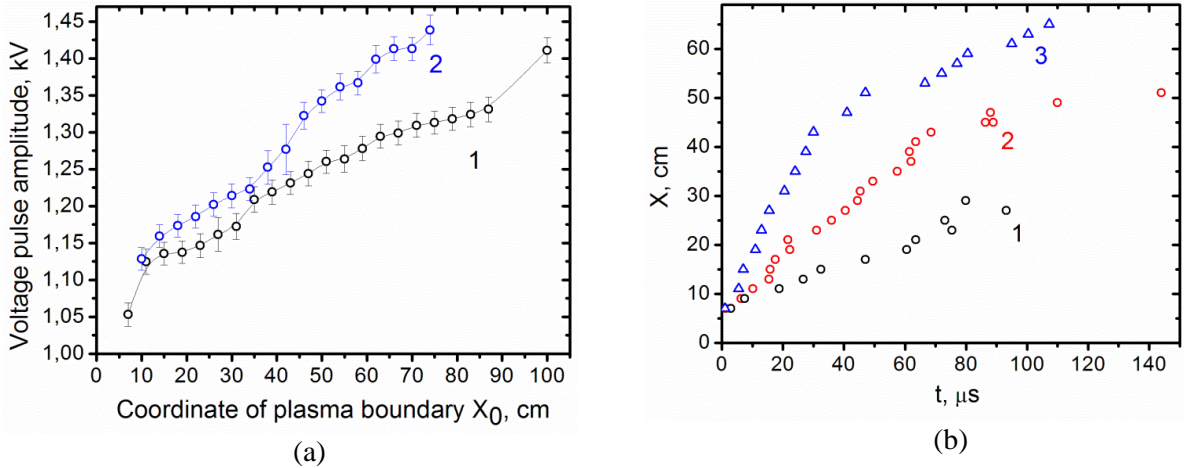

(a)      (b)

**Figure 4** a) Correlation between length of the OED in Ne and applied voltage for two pressures: 1 – 4 Torr, 2 – 1Torr; b) IW *x-t* diagrams for three lengths of one-electrode discharge: 1 – 25 cm, 2 – 51 cm, 3 – 70 cm

**2. Collision of two opposite ionization waves.** To apply the one-electrode discharge as a source of IW which has to be collided with we operated the discharge at voltages related to minimum IW attenuation, i.e. in the conditions when the discharge plasma can occupy the whole tube. Graphs on the figure 5 demonstrate *x-t* characteristics of two IW started from the opposite electrodes. Calculation of the *x* coordinate began from the right electrode, so the lower diagram corresponds to IW traveling from right hand side while the upper curve shows the motion of IW from the left side electrode.

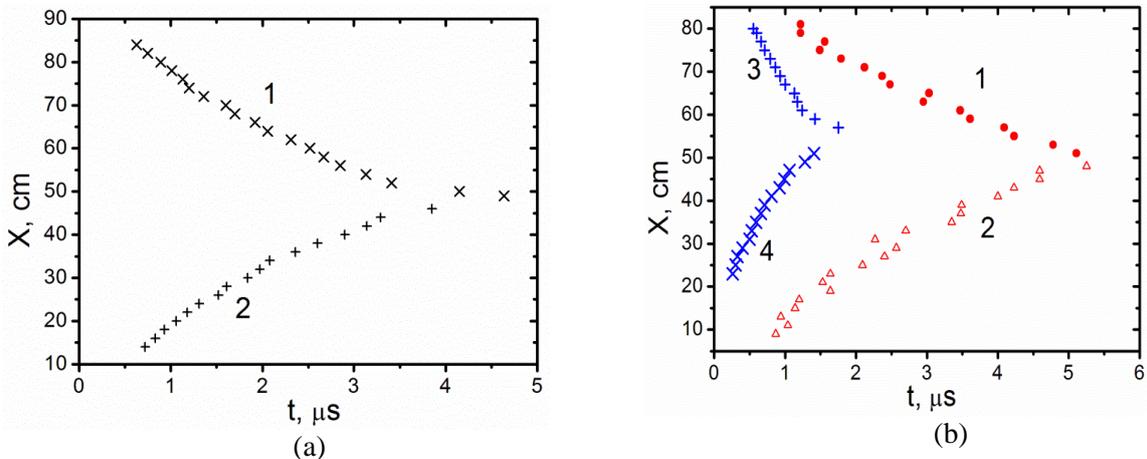

(a)      (b)

**Figure 5** *x-t* diagrams of two collided ionization waves started from: right electrode (2), (4); left electrode (1), (3). a) Ar $p = 4$ Torr, $U = 3.1$ kV, $f = 100$ Hz; b) Ne $p = 1$ Torr $U = 2.1$ kV, $f = 100$ Hz (curves 1 and 2); $U = 3.0$ kV, $f = 5$Hz (curves 3 and 4)

The curves demonstrate rapid deceleration (as compared with the case of only one wave) of the right IW which *x - t* characteristics transit to horizontal line showing the break of IW traveling at the middle of the tube. The deceleration of IW accompanied the decreasing intensity of the optical signal and widening of its pick. At the end point of the *x-t* characteristic the ionization wave attenuated so high that became undetectable. Along 1 –5 cm after the right IW quenching point the PMT did not register any emission signals (break between the diagrams).

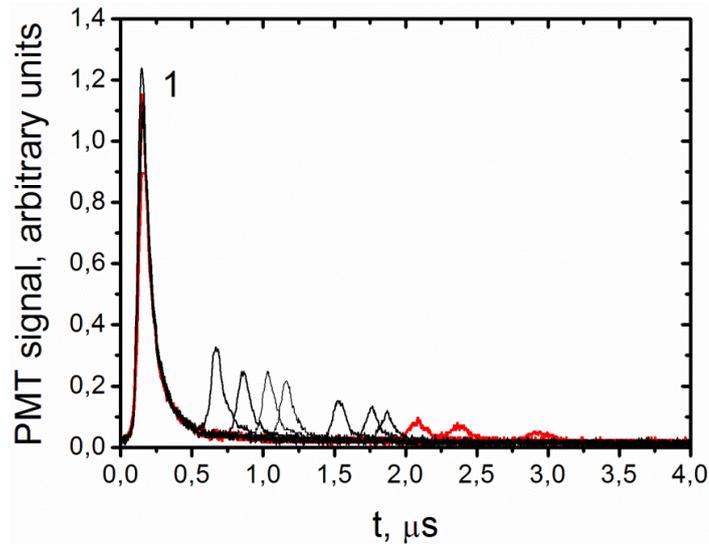

**Figure 6** Set of the right IW optical emission signals recorded in several *x* points with step 2 cm. Ne $p =$ 1Torr, $U =2.7$ kV, $f = 5$Hz

Series of the recorded optical signals oscillograms on the figure 6 depicts the situation described above for the right IW. The biggest maximum denoted as (1) corresponds to the fixed observation point near the electrode while the smaller-size maxima were obtained during the displacement of mobile fiber. At some distance ahead the collision area the ionization wave attenuates quickly and becomes low-observable then passing into the total suppression. Oscillograms corresponded to strongly attenuated IW are highlighted in red to be better distinguished against another maxima. At high rate of voltage pulses the suppression area is seen clearly (see photo on the figure 7) as a dark space against the background of the uniform glow of the discharge. In case of rather high voltages (over 3.0 kV) the suppression area is not fully dark: in its center a weakly glow plasma structure can be observed which is difficult to precisely detect. In coordinate points that are located after the suppression area in direction of the left electrode optical peaks from IW started to be detectable again. But now they are the signals of the left IW which motion characterizes the upper *x-t* diagrams at the figure 5. Indeed the left wave has the deceleration similar to the right one, but for clarity of presentation its kinematic characteristic is plotted together with the lower *x-t* diagram. This layout of the curves allows the ionization waves suppression to be easily shown and measured as a break space between the graphs. The table below contains the results of such measurements which show that suppression weakly extends with increasing voltage.

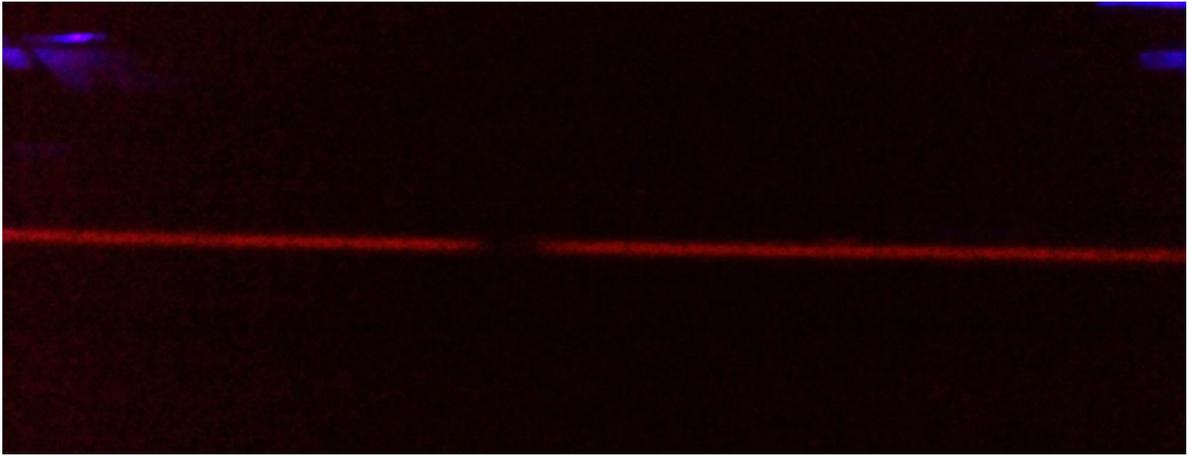

**Figure 7** Suppression of the opposite one-electrode discharges (the dark space in the discharge column). Ne $p = 1$ Torr, voltage pulses repetition rate is 100 Hz.

**Table 1** Sizes of the suppression area

| Gas | $p$, Torr | Voltage Amplitude, kV | Pulse repetition Rate, Hz | Size of suppression area, cm | $D$, cm |
|---|---|---|---|---|---|
| Ar | 4 | 3.1 | 100 | 3 | 42 |
| Ne | 1 | 2.1 | 5 | 1 | 10 |
| Ne | 1 | 2.7 | 5 | 5 | 20 |
| Ne | 1 | 3.0 | 5 | 5 | 38 |
| Ne | 1 | 2.1 | 100 | 2.5 | 11 |
| Ne | 1 | 3.0 | 100 | 5 | 38 |

Comparison (figure 8) of the *x-t* characteristics in case of only one wave and in case of two collided waves shows their marked difference. Curves (1) and (2) at the figure 8 as before depict the motion of two collided waves while the curve (3) corresponds to the case of only one IW traveled from the right electrode at the same conditions. As can be seen, the diagrams (1) and (3) plotted for the right IW run very close to each other up to approximately $x = 35$ cm. After this point the curve (1) rapidly declines toward the greater times which corresponds to the significant deceleration of the ionization wave before the collision which has to occur at point $x = 54$ cm. Therefore IW began to respond to the opposite wave at the 19 cm distance (for the example referred) before the collision event. The last column of the table 1 contains the distances $D$ between ionization waves at the points in which they begin to react to each other. Analysis of the table shows that the size of the IW suppression area as well as the distance $D$ strongly depend on the applied voltage amplitude but not on their repetition rate.

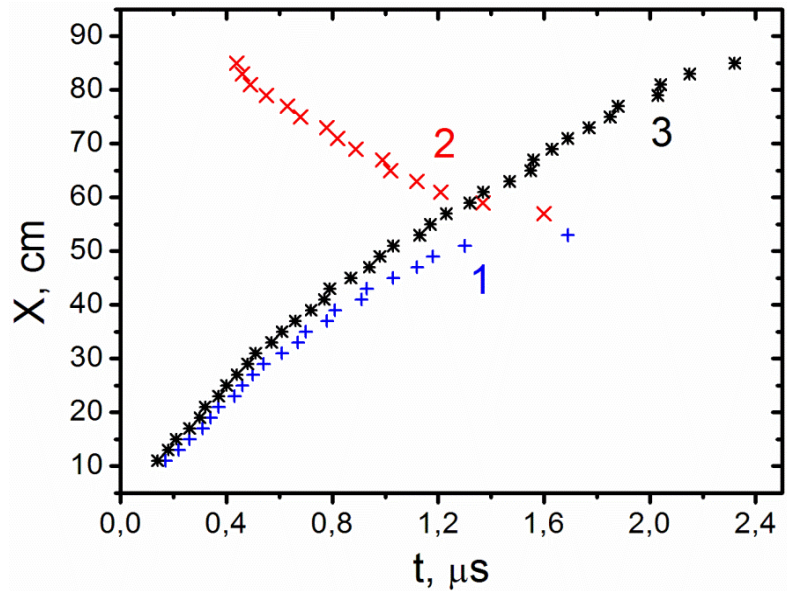

**Figure 8** Examples of *x-t* diagrams for collided IWs (1), (2) and one IW (3). Ne $p = 1$ Torr, $U = 3.0$ kV, $f = 100$ Hz

## IV. DISCUSSION

At the base of the *x-t* diagrams for the collided ionization waves it is possible to plot IW velocity as a function of *x* coordinate. In contrast to the case of *x-t* characteristics for one IW which are close to the strength line (weak attenuation of IW) the cases of the collided waves are characterized by their appreciable attenuation. This fact permits to make the simplifying assumption that IW has the constant deceleration and fits the *x-t* characteristics with second order curve. Plots in the figure 9 show the result of such calculation for 4 Torr argon: red dots – right IW, blue dots – left IW. Plots for the collided waves accompanied with line (3) which corresponds to the propagation of one IW from the right electrode. The approximation gives constant negative accelerations for both opposite waves: -2.1 cm/(μs)$^2$ for the right IW and -2.0cm/(μs)$^2$ for the left one.

If we consider two kinematic characteristics for the right IW in case of a single wave (curve 3) and in case of a collided wave (curve 1) at small distances from the electrode (x ≤ 30cm), we will see that both plots slightly differ from each other. The wave to be collided has a lower speed than that of the single wave, and this difference linearly increases in time with rate $4.1 \cdot 10^6$ cm/s$^2$. This means that despite the same potentials of electric field at the electrodes the interacted waves develop with a lower speed from the breakdown beginning. That effect can be attributed with lower current $i_p$ beyond IW front in case of opposite waves. To ensure the development of two IW similar to that in case of one active electrode the external circuit must carry high current.

According to the accepted models, examples of which can be found in papers [6, 22, 23] the ionization wave propagation can be briefly described as follows. Movement of IW is accompanied with charged particles production ahead the front. In case of positive polarity of voltage the electrons flow into the front and then to the external circuit leaving the positive space charge which forms the ionization front in the next point. Closing of the electron current occurs via displacement current which is attributed with charging of the tube distributed capacity. As a result production of the space charge is inevitably accompanied with charging of the tube wall. The rate of this process which is characterized by the charging current $i_c$ should determine the ionization wave velocity.

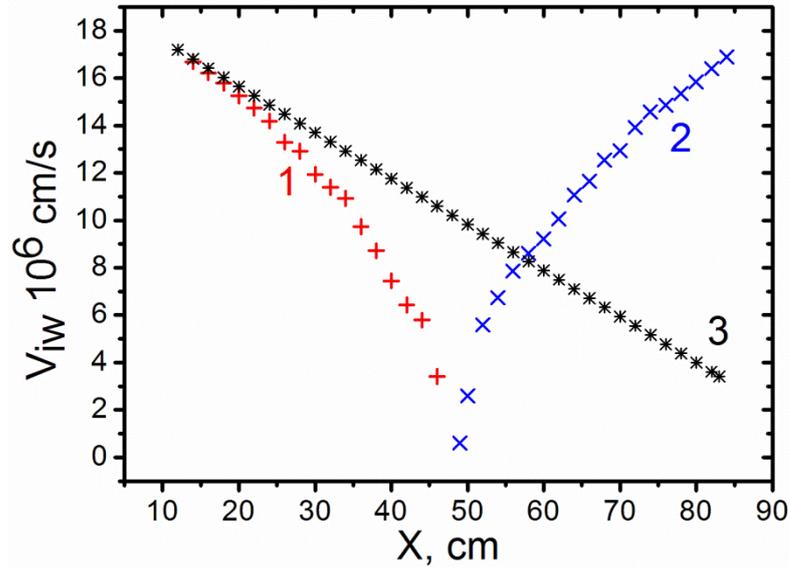

**Figure 9** IW velocity $V_{iw}$ as a function of coordinate $x$, (1) right IW; (2) left IW; (3) single IW traveled from the right electrode. Ar $p$ = 4 Torr, $U$ = 3.1 kV, $f$ = 100 Hz

The charging current depends on the value of distributed capacity and on the relation between impedance of external circuit Ze and impedance of plasma Zp beyond the front and ip = ic. Value of Ze restricts current in the circuit, thus the development of single IW will take the same current ip1 as in case of collided waves for the same voltage amplitude. The crude estimation of the discharge currents amplitudes in both cases gives ip2/ip1≈1/2. In other words, in experimental scheme used the total current of single IW must be equal to the sum of currents of two opposite waves. Therefore the value of the current does not influence the IW velocity drastically compared to the voltage amplitude.

At greater distances from the electrode ($x \geq 30$cm) the *x-t* characteristics diverge substantially; the collided waves attenuate rapidly and pass into the suppression region. The suppression phenomenon can be illustrated on the base of a simplified qualitative electrostatic model. In the model the ionization front consisted of a space charge and a current channel between it and the active electrode simulated as a conducted bar with resistance $R_p$ having charged semispherical head. Such approach is executed for estimation of electric field distribution over leader of lightning [24]. Since the IW plasma spreads over the whole tube cross-section we choose the bar radius being equal to the tube radius $r_0$. The same radius is chosen for spherical end of the bar which simulates elongated shape of positive IW front generally observed in experiments [22, 25]. The bar transmits the electrode potential into the tube volume minus voltage drop at $R_p$; in simplest consideration this difference may be neglected. Maximum electric field $E$ near the semi-sphere surface determines the rate of ionization before IW front and therefore its velocity; if $\varphi_0$ is the electrode potential transmitting by the wave into the tube volume then: $E \approx \varphi_0/2r_0$ [24]. The charge of the semi-spherical head will be $Q = 2\pi\varepsilon_0 r_0 \varphi_0 \approx 2$nC for $\varphi_0 = 3.0$ kV and $r_0 = 1$ cm which qualitatively matches the results of the paper [11].

The collision of two ionization waves can be modeled as approaching of the two bars assuming that the rate of this motion strongly depends on axial electric field in space between them (figure 10). If the resultant electric field at some point $x_0$ is $E(x_0) \leq E_i$ where $E_i$ is the ionization threshold the IW cannot approach $x_0$. Numerous data of measurements of the electric field in the front of IW give typical value of $E/N \sim 10^3$ Td [11, 22, 27] which corresponds to $E \approx 3\cdot 10^2$ V/cm at gas pressure 1Torr. If we take this value as realistic evaluation for $E_i$ the model gives for $x_0 \approx (2/3) \cdot \sqrt{r_0 \varphi_0 / E_i}$ therefore $x_0 \approx 2$ cm. It is obviously that the suppression area appears due to the weak electric field between two equally charged fronts thus we can assume that its length is $L = 2x_0$ which is close to the observed values.

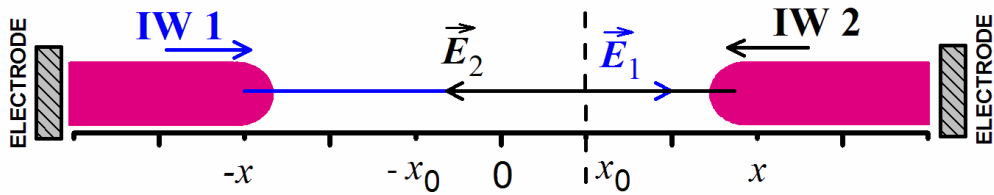

**Figure 10** Schematic picture illustrating collision of two positive ionization waves.

The toy model discussed above leads to the conclusion that electrostatic interaction between IW becomes significant at distances about $2x_0 \approx 4r_0$ but deceleration of the waves begins at distances up to $30r_0$. Such behavior is unexpected under the assumption about electrostatic type of the electric field in IW front. This field produces ionizing multiplication of initial electrons ahead the front which are generated by photoionization caused the IW ultraviolet radiation. Important aspect is the interaction of IW plasma with the tube wall which is manifested in that the current flowing in the front charges the wall surface up to the potential closing to that in IW. In models [2, 6, 23] this process in a screened tube is equivalent to the motion of voltage wave in a long transmission line with constant distributed capacity $C$ so that the discharge is replaced with the simplifying infinite chain from resistors and capacitors. In reality the current $i_c$ distributes over the wall section which is nearest to the IW front and depends on a distance from the ionization front and is looped with displacement current. In case the tube is not covered with metal screen the displacement current closes at surrounding elements of the set-up. The field of displacement current lines spreads over the space almost uniformly around the IW front at a distance $d$ equaling approximately the length to the nearest grounded surface. When the opposite source with the same field of the displacement current occurs at the distance $D \leq 2d$ the current lines fluxes subtract from each other thus decreasing each of them. This leads to a decrease of total displacement current as well as of the wall charging current, and therefore to lowering the rate of IW front development. Accommodation of the tube ensures value of $d$ not less than 20 cm, so the maximum $D$ is about 40 cm. The data in Table1 generally confirms this supposition.

## V. CONCLUSIONS

The case of ionization waves interaction being their mutual suppression in long tubes at low gas pressure was observed. Necessary condition of the suppression is the opposite traveling of two independent ionization waves with the same polarity triggered by the voltage pulses with the same amplitudes. The phenomenon was found in neon and argon via simultaneous generation of two opposite one-electrode discharges which represented the multiple repeated ionization waves generation at the tube electrodes. The kinematic characteristics of IW of both unilateral and bilateral one-electrode discharge were recorded and compared. It was shown that two opposite IW rapidly attenuates in contrast with case of a single wave. The attenuation has the highest degree close to the center of the tube, and results in the existence of 1 – 5 cm wide area in which the ionization wave signals are not registered. At the same time the zone at the *x-t* diagrams of rapid deceleration of IW begins at a 20 cm distance ahead of the suppression achievement. Several qualitative estimations and analyses were presented.

**Acknowledgments**

This work was supported by the Russian Foundation for Basic Research (grant No18-32-00223).